\documentclass[conference]{IEEEtran}
\usepackage[letterpaper, left=0.625in, right=0.625in, bottom=1in, top=0.75in]{geometry}
\IEEEoverridecommandlockouts
\usepackage{cite}
\usepackage{amsmath,amssymb,amsfonts}
\usepackage{graphicx}
\usepackage{textcomp}
\usepackage{subcaption}
\usepackage{xcolor}
\usepackage{algorithm}
\usepackage{algpseudocode}
\usepackage{xltabular}
\usepackage{comment}
\usepackage{tabularx}
\usepackage{caption}
\usepackage{multicol}
\usepackage{float}
\usepackage{ amssymb }
\usepackage{lipsum}
\usepackage{multirow}
\usepackage{array}
\usepackage{mathtools}
\usepackage{environ}         
\usepackage{etoolbox}        
\usepackage{graphicx}
\newlength{\myl}
\let\origequation=\equation
\let\origendequation=\endequation
\RenewEnviron{equation}{
	\settowidth{\myl}{$\BODY$}                       
	\origequation
	\ifdimcomp{\the\linewidth}{>}{\the\myl}
	{\ensuremath{\BODY}}                             
	{\resizebox{\linewidth}{!}{\ensuremath{\BODY}}}  
	\origendequation
}
\def\BibTeX{{\rm B\kern-.05em{\sc i\kern-.025em b}\kern-.08em
    T\kern-.1667em\lower.7ex\hbox{E}\kern-.125emX}}
\begin{document}
\title{Blue Data Computation Maximization in 6G Space-Air-Sea Non-Terrestrial Networks}
\author{\IEEEauthorblockN{Sheikh Salman Hassan$^1$, Yan Kyaw Tun$^1$, Walid Saad$^2$, Zhu Han$^3$, and Choong Seon Hong$^1$}
\IEEEauthorblockA{$^1$Department of Computer Science and Engineering, Kyung Hee University, Yongin, 17104, Republic of Korea\\
$^2$Wireless@VT, Bradley Department of Electrical and Computer Engineering, Virginia Tech, VA, 24061, USA\\
$^3$Department of Electrical and Computer Engineering, University of Houston, Houston, TX 77004-4005 USA}
Email: \{salman0335, ykyawtun7, cshong\}@khu.ac.kr, walids@vt.edu, hanzhu22@gmail.com.
\thanks{This work was partially supported by the National Research Foundation of Korea (NRF) grant funded by the Korea government (MSIT) (No. 2020R1A4A1018607) and by the Institute of Information \& communications Technology Planning \& Evaluation (IITP) grant funded by the Korea government (MSIT) (No.2019-0-01287, Evolvable Deep Learning Model Generation Platform for Edge Computing) *Dr. CS Hong is the corresponding author.}}
\maketitle
\begin{abstract}
Non-terrestrial networks (NTN), encompassing space and air platforms, are a key component of the upcoming sixth-generation (6G) cellular network. Meanwhile, maritime network traffic has grown significantly in recent years due to sea transportation used for national defense, research, recreational activities, domestic and international trade. In this paper, the seamless and reliable demand for communication and computation in maritime wireless networks is investigated. Two types of marine user equipment (UEs), i.e., low-antenna gain and high-antenna gain UEs, are considered. A joint task computation and time allocation problem for weighted sum-rate maximization is formulated as mixed-integer linear programming (MILP). The goal is to design an algorithm that enables the network to efficiently provide backhaul resources to an unmanned aerial vehicle (UAV) and offload HUEs tasks to LEO satellite for blue data (i.e., marine user's data). To solve this MILP, a solution based on the Bender and primal decomposition is proposed. The Bender decomposes MILP into the master problem for binary task decision and subproblem for continuous-time resource allocation. Moreover, primal decomposition deals with a coupling constraint in the subproblem. Finally, numerical results demonstrate that the proposed algorithm provides the maritime UEs coverage demand in polynomial time computational complexity and achieves a near-optimal solution. \\
\end{abstract}
\section{Introduction}
Non-terrestrial networks (NTN), composed of space and airborne platforms, will be a key component of the upcoming sixth-generation (6G) wireless cellular systems as a means to provide the ubiquitous and ultra-high capacity wireless connectivity \cite{walid6G}. As a supplement to terrestrial infrastructure, space, and airborne stations have enormous potential for promoting flexible global connectivity in densely populated areas, cost-effective network coverage in public safety situations, last-mile service delivery, and backhaul in remote, rural, and difficult-to-access zones \cite{6G_SURVEY}. 

Meanwhile, maritime transportation is rapidly emerging as a significant component of our transportation infrastructure for domestic and international trade, national security, scientific research, aquaculture (fastest-growing food sector), and recreational activities \cite{literature1}. The maritime world faces many challenges \cite{maritime_chal}. Therefore, the United Nations coined the term \emph{blue economy} to improve human well-being and social equity while significantly reducing environmental risks and ecological scarcities related to oceans. Growing global trade requires more and more ships, many of which have to safely and productively sail international waters and harbors. Reliable maritime transportation depends on seamless connectivity with terrestrial networks to ensure accurate and intelligent navigational communications. Similarly, the maritime internet-of-things (M-IoT) applications are rapidly increasing \cite{MIOT}. Therefore, NTN is a suitable candidate for reliable and seamless maritime network connectivity.

Based on the preceding discussion, we propose to use jointly the low earth orbit (LEO) satellite and unmanned aerial vehicle (UAV) empowered with multi-access edge computing (MEC) for maritime users \cite{ICC_maritime}, \cite{time-sensitive-iot}. Due to a scarcity of optical fibers and base stations, maritime communications must operate in a highly complex and heterogeneous setting, thus posing significant challenges to reliable transmission and traffic steering performance for service-oriented maritime communication networks. Existing work more focused on marine users association and communication \cite{coordinated_maritime}, \cite{ownicoin2021}.  

The main contribution of this paper is to fill the knowledge gap of marine user's task computation mechanism. In particular, we consider a heterogeneous 6G space-air-sea non-terrestrial network (SAS-NTN) to meet the increasing maritime network requirements. In this system, an LEO-MEC satellite and a UAV-MEC will aid existing offshore base stations, which work in a coordinated fashion for better maritime coverage. LEO-MEC and UAV-MEC are proposed as a seamless and reliable method to enhance the computing capability of high (HUE) and low-antenna gain user equipments (LUE) respectively. LEO-MEC allows the HUEs to offload their intensive \emph{blue data (generated by maritime users)} 
computations task in a binary offloading manner. Our key contributions can thus be summarized as follows:
\begin{itemize}
\item We propose an LEO-MEC satellite and UAV-MEC-based 6G SAS-NTN architecture by considering both variants of maritime users, i.e., high and low antenna gain users.
\item We propose a binary task computation policy for HUEs depending upon their channel condition.
\item We develop an iterative algorithm based on the Bender and primal decomposition to solve the proposed mixed-integer linear programming (MILP), demonstrating the effectiveness in the simulation results.
\end{itemize}
\section{System Model and Problem Formulation}
\label{system_model}
\subsection{Network Model}
\begin{figure}[t]
\centering
{\includegraphics[width=\columnwidth, height=2.18in]{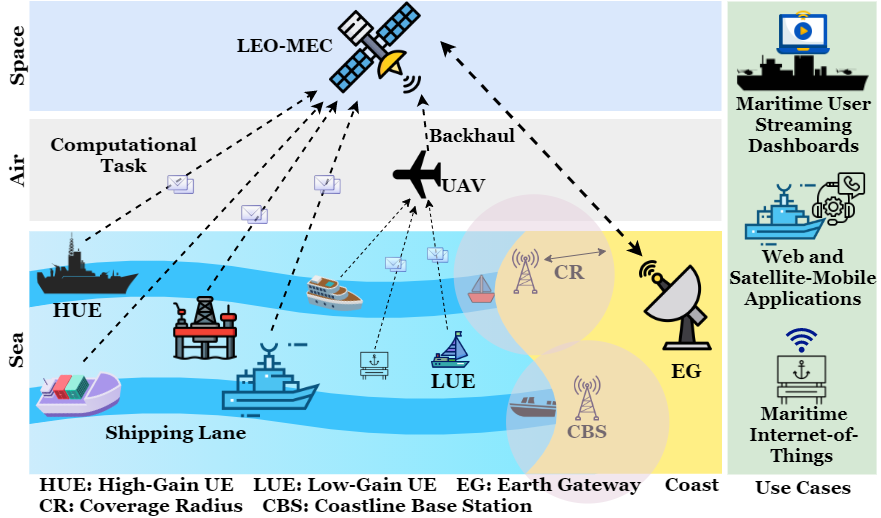}
\caption{Illustration of LEO-MEC empowered 6G SAS-NTN.}
\label{sysmodel}}
\vspace{-0.2in}
\end{figure}

As shown in Fig. \ref{sysmodel}, we consider a heterogeneous 6G SAS-NTN maritime communication network. SAS-NTN consists of MEC empowered LEO satellite\footnote{Hereinafter, the LEO-MEC is considered as an LEO-MEC satellite unless otherwise stated.}, and a flying UAV-MEC serves as an aerial base station. We define a set $\mathcal{C}$ of coastal base stations (CBS), having a limited coverage radius that cannot satisfy the deep-sea maritime users' demand. Consequently, the LEO-MEC, a suitable candidate for service provisioning to these UEs, is considered. The LEO-MEC has considered a single high-gain antenna that can service HUEs task computation and UAV-MEC backhauling. Moreover, maritime UEs can be of two types: a set $\mathcal{M}_h$ of $M_h$ HUEs and a set $\mathcal{M}_l$ of $M_l$ LUEs. HUE and LUE can be differentiated based on the power of their antenna gain. Specifically, we consider network resource management for HUEs and UAV-MEC backhauling. Additionally, we assume HUEs are deployed on their specific pre-defined shipping routes.
\subsection{Channel Model}
We assume that the HUEs and the UAV-MEC need to send their data to LEO-MEC in the studied time of duration $T$. Moreover, each timeslot of $T$ can be allocated to each HUE depending upon their channel condition and UAV-MEC. The LEO-MEC communicates with HUEs and the UAV-MEC in a time division multiple access (TDMA) fashion to avoid network interference. Both communications links in the scenario operate on the Ka-band (30 GHz), which is a well-known millimeter wave (mmW) carrier frequency for satellite communication as discussed in \cite{mm_wave_walid}. The TDMA scheme is used to enable directional transmissions in the considerably high path loss mmW band while maintaining low complex designs for the very-small-aperture terminal (VSAT). We consider the HUEs to be equipped with a single antenna. Therefore, the composite channel capture line-of-sight (LoS) and non-line-of-sight (NLoS) gain between LEO-MEC and its associated HUE $i$ can be defined as:
\begin{equation}
\begin{aligned}
    g_i =  \alpha10^{\frac{-\beta_i}{10}} G_s G_i, \quad \forall i \in \mathcal{M}_h,
\end{aligned}
\end{equation}
where $\alpha$ denotes the Rician fading channel coefficient, $G_s$ and $G_i$ denote the antenna gain of LEO-MEC and HUE respectively, and $\beta_i = \Tilde{\gamma} + 10\gamma\log_{10}(\frac{d_i}{d_0})+\phi$ captures the large-scale fading on mmWave between LEO-MEC and HUE $i$. Here, $\Tilde{\gamma}$ represents the intercept parameter (path loss at reference distance $d_0$), ${\gamma}$ is the slope of the fit (path loss exponent), $d_i$ reflects this distance between LEO-MEC and HUE $i$, and $\phi$ is the model deviation in fitting represented by a zero-mean Gaussian random variable with standard deviation $\omega$ \cite{mm_wave_walid}. Similarly, the channel gain between LEO-MEC and UAV can be defined as:
\begin{equation}
\begin{aligned}
     g_u =  \alpha10^{\frac{-\beta_u}{10}} G_s G_u,
\end{aligned}     
\end{equation}
where $G_u$ is the UAV antenna gain and $\beta_u$ is large-scale channel fading between LEO-MEC and UAV-MEC.
\subsection{Local Computation Model}
The HUEs utilize store energy in local task computational mode. Each HUE $i$ has a process computing speed $f_i$ (Hertz) and and a computation time $\tau_i$. Therefore, each HUE can process $f_i \tau_i/\chi$ bits, where $\chi$ is the required number of cycles to process one bit of computational task data. Additionally, each HUE has an energy budget constrained by $\nu_i f_i^3 \tau_i \leq E_{\mathrm{th}}$, where $\nu_i$ represents the energy efficiency coefficient of processor chip \cite{DROO}. The HUEs local computation can be maximized under the studied time and energy budget. Each HUE can compute the task throughout the time, i.e., $\tau_i^* = T$ and  process computing speed will be $f_i^* = \big( \frac{E_{\mathrm{th}}}{\nu_iT} \big)^{\frac{1}{3}}$. Hence, the HUEs local computational rate (bits/s) will be:
\begin{equation}
\begin{aligned}
    R^\mathrm{Local}_i(\tau_i) =  \frac{f_i^* \tau_i^* }{\chi T}, \quad \forall i \in \mathcal{M}_h.
\end{aligned}
\end{equation}
\subsection{UAV-MEC Backhaul and LEO-MEC Computation Model}
\begin{figure}[t]
\centering
{\includegraphics[width=\columnwidth]{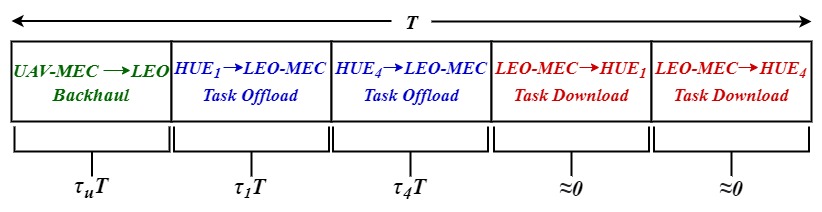}
\caption{Example of LEO-MEC time resource allocation.}
\label{timeframe}}
\vspace{-0.2in}
\end{figure}

It is assumed that UAV-MEC and HUEs cannot transmit simultaneously. Therefore, firstly the UAV-MEC can use backhaul resource for the duration of $\tau_u \in [0,1]$ and secondly, associated HUEs will offload their task according to their received time duration, i.e., $\tau_i \in [0,1]$ in each timeslot as illustrated in Fig. \ref{timeframe}. We assume that the LEO-MEC has significant energy resources and computational power. Therefore, the time of task computation at LEO-MEC and feedback of computation results to UAV and HUEs can be neglected \cite{result_neglect}. Hence, the studied time can be divided into UAV-MEC backhaul resources and HUEs task offloading as:
\begin{equation}
\begin{aligned}
\tau^{\mathrm{UAV}}_u + \sum_{i=1}^{M_h} \tau^{\mathrm{HUE}}_i = T.
\end{aligned}
\end{equation}
The UAV-MEC backhaul throughput in each timeslot can be define as:
\begin{equation} \label{rate_uav}
\begin{aligned}
    R^{\mathrm{UAV}}(\tau_u) = \tau_u\frac{B}{\mu_u}\log_2 \Big( 1 + \frac{g_u P_u}{\sigma^2} \Big),
\end{aligned}
\end{equation}
where $\tau_u$ is the time duration for UAV backhauling, $B$ is the network bandwidth, $\mu_i$ is the communication overhead for task offloading, e.g., packet header and encryption \cite{FOLLOW2}, $P_u$ is the UAV transmit power, and $\sigma^2$ is the LEO-MEC received noise power. Additionally, each HUE $i$ transmit a power, i.e., $P_i = \frac{E_{\mathrm{th}}}{\tau_i}$ to offload the task. Thus, the computational rate which is equal to the throughput capacity of HUE $i$ can be defined as:
\begin{equation} \label{rate_leo}
\begin{aligned}  
    R^\mathrm{LEO}_i(\tau_i) = \tau_i\frac{B}{\mu_i}\log_2 \Big( 1 + \frac{g_i P_i}{\sigma^2} \Big), \quad \forall i \in \mathcal{M}_h,\\ 
    = (T-\tau_u)\frac{B}{\mu_i}\log_2 \Big( 1 + \frac{g_iP_i}{\sigma^2} \Big), \quad \forall i \in \mathcal{M}_h,
\end{aligned}
\end{equation}
where $\tau_i$ is the time duration allocates to HUE $i$ and $P_i$ is the transmit power of each HUE $i$. Both (\ref{rate_uav}) and (\ref{rate_leo}) are increasing function of time.
\subsection{Problem Formulation}
\label{problem_formulation}
The channel gain $g = \{g_u,~g_i|i \in  \mathcal{M}_h\}$ is a time-varying parameter in the studied period, while the remaining parameters are constant. Thus, the association of HUE $i$ depends upon the channel gain $g$ in the studied time. The weighted computational and communication sum-rate of the LEO-MEC for maritime network in the studied time can be given by:
\begin{equation}
\begin{aligned}
R(\boldsymbol{g}, \boldsymbol{y}, \boldsymbol{\tau}_i, \tau_u) \triangleq \sum_{i=1}^{M_h} z_i \Big( (1-y_i)R^{\mathrm{Local}}_{i} + y_i R^{\mathrm{LEO}}_{i} \Big) + R^{\mathrm{UAV}},
\end{aligned}
\end{equation}
where $z_i$ is unique weight for each HUE which depends upon their contribution in computation, $y = \{ y_i|i \in \mathcal{M}_h\}$ is the HUEs offloading decision variable, and $\tau = \{ \tau_i|i \in \mathcal{M}_h\}$ is the duration of allocated time to each HUE $i$. Here, $y_i \in \{0,1\}$, where $y_i = 1$ indicates that HUE $i$ will transmit its computation task to the LEO-MEC and $y_i = 0$ implies a local task computation. Our objective is to maximize the weighted sum of computational and communication rate in studied time with channel realization $\boldsymbol{g}$. Therefore, we can define the optimization problem as:
\begin{subequations} \label{optimization1}
\begin{flalign}
R^*(\boldsymbol{g})=\underset{\boldsymbol{y},\tau_u, \boldsymbol{\tau}_i}{\text{max}} \quad &  R(\boldsymbol{g}, \boldsymbol{y}, \boldsymbol{\tau}_i, \tau_u), \label{objective1}\\
\text{s.t.}  \quad   &     \tau_u + \sum_{i=1}^{M_h} \tau_i \leq T,    \label{eq:constraintp1_1}\\
&       \tau_u > 0, \tau_i \geq 0, \quad \forall{i} \in  \mathcal{M}_h, \label{eq:constraintp1_2}\\
&       y_i \in \{0,1\}, \quad \forall{i} \in   \mathcal{M}_h. \label{eq:constraintp1_3}
\end{flalign}
\end{subequations}
The proposed problem (\ref{optimization1}) is an MILP, which is a non-deterministic polynomial-time (NP-hard) problem. Although, if $\boldsymbol{y}$ is given, this problem can be modified as linear programming which is describe in Section \ref{Proposed_algorithm}.
\section{Proposed Algorithm}
\label{Proposed_algorithm}
In this section, we propose the joint Bender and primal decomposition to deal with the problem (\ref{optimization1}). The major challenge of solving (\ref{optimization1}) lies in the task computation decision problem. Hence, we provide Bender decomposition (BD) to decompose the main problem (\ref{optimization1}) into two subproblems, i.e., task computation decision and time resource allocation problem. Furthermore, the time resource allocation problem will be handled by primal decomposition.
\subsection{Bender Decomposition}
BD is a well-known solution technique to solve MILP problem \cite{conejo2006decomposition}. This algorithm decomposes the MILP into binary variables integer programming as a master problem and continuous variables linear programming as a subproblem \cite{ownicoin2021}. The main goal of the master problem is to solve an integer linear programming problem, while the subproblem is used to solve a continuous linear programming problem. This algorithm uses an iterative method that can reach a point of convergence when the pre-defined criterion on upper and lower bounds is met. 

The initialization process happens first with a particular scenario. By fixing integer variables, the subproblem is solved by using only continuous variables. The subproblem will provide the optimal value of continuous variables (time allocation). After getting this solution, we produce the Bender cut for feasibility and optimality conditions by getting dual variables associated with fixed value integer variables, which allows us to solve the master problem. As a result, the disparity between the upper and lower bounds serves as the stopping condition for this algorithm. The joint MUEs task decision and time resource allocation problem-solution method will follow the steps described next.\\
\subsubsection{Initialization}
We consider that the master problem can generate trivial solutions, and therefore, arbitrary task decisions are initially generated. Subsequently, the loop counter is initialized, i.e., $j=1$. The task offloading decision variable is a binary indicator and, thus, their upper and lower bound can be initialized as $y_{\mathrm{UB}} =1$ and $y_{\mathrm{LB}}=0$, respectively. Moreover, we use an auxiliary variable in the form of a function $\Psi$, representing the objective function value of subproblem in each iteration $j$ into the master problem objective function. In first iteration $j=1$, the function $\Psi^{j}$ can be initialized as $\Psi^{\textrm{down}}$, to restrict the subproblem from infeasible solution.\\
\subsubsection{Subproblem}
We assume that the channel realization $\boldsymbol{g}$ in each iteration $j$ is obtained as $\boldsymbol{g}^j$. Thus, the optimal solution of time resource allocation from linear programming, i.e., $\{\tau^*_u, \boldsymbol{\tau}^*_i \}$ can be found by a low-complexity algorithm that is suitable for dynamic SAS-NTN environment. The subproblem can be defined as: 
\begin{subequations} \label{optimization3}
\begin{flalign}
\underset{\boldsymbol{y}, \tau_u, \boldsymbol{\tau}_i}{\text{max}} \quad &  R(\boldsymbol{{y}}^{*j}, \tau_u, \boldsymbol{\tau}_i), \label{objective3}\\
\text{s.t.}  \quad   &     \tau_u + \sum_{i=1}^{M_h} \tau_i \leq T,    \label{eq:constraintp3_1}\\
&       \tau_u > 0, \tau_i \geq 0, \quad \forall{i} \in   \mathcal{M}_h, \label{eq:constraintp3_2}\\
&       y_{i} = y_{i}^{*j}: ~\kappa_{i}^j, \quad \forall{i} \in   \mathcal{M}_h, \label{eq:constraintp3_3}
\end{flalign}
\end{subequations}
where $\boldsymbol{{y}}^{*j}$ is the optimal task decision in iteration $j=1$ and $\kappa_{i}^j$ represents the dual variable value associate with constraint (\ref{eq:constraintp3_3}) of fixed task decision variable $y_{i}^{*j}$ in each iteration $j$.\\
\subsubsection{Bounds Calculation \& Convergence Analysis} The convergence of the BD can be analyzed by taking the difference between upper and lower bounds. The lower bound is given by computing the optimal value of an objective function in problem (\ref{Master_problem}) for each iteration $j$, which can be defined as:
\begin{equation}
\begin{aligned}
    R_{\textrm{LB}}^j =  \Psi^{j},  \label{lower_bound}
\end{aligned}
\end{equation}
where $R_{\textrm{LB}}^j$ progressively less and less relaxed as iteration increase, therefore, lower bound is monotonously increasing with the iterations \cite{conejo2006decomposition}. Subsequently, the upper bound is given by computing the optimal value of objective function in problem (\ref{optimization1}) for each iteration $j$ as:
\begin{equation}
\begin{aligned}
    R_{\textrm{UB}}^j = R(\boldsymbol{\Tilde{y}}^j, \tau_u^j, \boldsymbol{\tau}_i^j). \label{upper_bound} 
\end{aligned}
\end{equation}
After acquiring both bounds, the difference between them will decide the convergence and stopping criterion which can be defined as, 
 \begin{equation}
 		\begin{cases}
 			R_{\textrm{UB}}^j - R_{\textrm{LB}}^j \leq \epsilon, & \text{stop},\\
 			\textrm{otherwise},                                           &\text{continue},\\
 		\end{cases}
\end{equation}
and the tolerance parameter $\epsilon$ is pre-defined. Following convergence, the optimal value of $\{\tau^*_u, \boldsymbol{\tau}^*_i\}$, and $\boldsymbol{y}^*$ can be achieved.\\
\subsubsection{Master Problem}
The master problem deals with binary task decisions in each iteration $j$. After the increment in loop counter, i.e., $j=j+1$, the master problem solves which can be define as:
\begin{subequations} 
\label{Master_problem}
\begin{align}
\max_{\boldsymbol{y}, \Psi} \quad   & \Psi,  \label{MP_objective}\\
\text{s.t.}  \quad                  &  \Psi \leq \sum_{i=1}^{M_h}{\kappa}_{i}^{k}({{y}}_{i}-{{y}}^{*k}_{i})+R(\boldsymbol{y}^{*k}_i,{\tau}^{*k}_u, \boldsymbol{\tau}^{*k}_i), \nonumber \\ 
                                    & \qquad k=1,...,j-1, \label{MP_constr1}\\ 
			                        & \Psi \geq \Psi^{down}, \label{MP_constr2}\\
			                        & \sum_{i\in\mathcal{M}_h } y_{i} \le 1, \label{MP_constr3}\\
			                        &  y_{i} \in \{0,1\}, \quad \forall i \in \mathcal{M}_h,   \label{MP_constr4}
\end{align}
\end{subequations}
where the inequality constraints in (\ref{MP_constr1}) and (\ref{MP_constr2}) indicate the Bender optimality and feasibility cuts, respectively. In each iteration, a new Bender cut is added to the master problem. The Bender cuts approximate the objective function of the subproblem (\ref{optimization3}) from below, which compose of $j-1$ hyperplanes. The details of the BD are given in Algorithm \ref{BD}.
\begin{algorithm}[t]
		\caption{Data Computation Maximization and Task Decisions by Bender Decomposition} \label{BD}
		\begin{algorithmic}[1]
			\State Initialize: loop counter $j=1$, $\Psi^{\mathrm{down}}$, $\epsilon$,
			\While {$R_{\textrm{UB}}^j - R_{\textrm{LB}}^j > \epsilon$}
			\State\textbf{Subproblem}
			\State \hskip1.5em Compute optimal ${{\tau}^{*j}_{u}}$, $\boldsymbol{{\tau}^{*j}_{i}}$, and $\kappa_{i}^j$ by Algorithm \ref{PD}
			\State \textbf{Convergence Analysis}
			\State \hskip1.5em Compute the lower ($R_{\textrm{LB}}^j$) and upper ($R_{\textrm{UB}}^j$) bounds \indent \quad by (\ref{lower_bound}) and (\ref{upper_bound})
			\State\textbf{Master Problem}
			\State \hskip1.5em Step 1: Update the loop counter $j=j+1$
			\State \hskip1.5em Step 2: Add new cut to the Master problem (\ref{Master_problem})
			\State \hskip1.5em Step 3: Solve the updated master problem
			\State \hskip1.5em Step 4: Compute the optimal value of $\boldsymbol{{y}^{*j}_{i}}$ and $\Psi^j$
			\EndWhile    
		\end{algorithmic}
	\end{algorithm}
\subsection{Primal Decomposition} To deal with the time resource coupling constraint in (\ref{eq:constraintp3_1}), we apply the primal decomposition. We use transformation on this constraint, and convert it into coupling variable . Thus, an auxiliary variable $\theta$ is introduced and can be given as \cite{boyd2004convex}:
\begin{align}
      \tau_u \leq \theta, \\
    \sum_{i=1}^{M_h} \tau_i \leq T-\theta.
\end{align}
At this point, our subproblem can be modified as:
\begin{subequations} 
\label{SP_mod}
\begin{flalign}
\underset{\boldsymbol{y}, \tau_u, \boldsymbol{\tau}_i}{\text{max}} \quad &  R(\boldsymbol{{y}}^{*j}, \tau_u, \boldsymbol{\tau}_i), \label{SP_OBJ_MOD}\\
\text{s.t.}  \quad   &  \tau_u \leq \theta,  \label{SP_MOD_C1} \\
                     &   \sum_{i=1}^{M_h} \tau_i \leq T-\theta, \label{SP_MOD_C2}\\  
                     &       \tau_u > 0, \quad \tau_i \geq 0, \quad \forall{i} \in   \mathcal{M}_h,\label{SP_MOD_C3} \\
                     &       y_{i} = y_{i}^{*j}:~\kappa_{i}^j, \quad \forall{i} \in   \mathcal{M}_h. \label{SP_MOD_C4} 
\end{flalign}
\end{subequations}
By fixing the value of $\theta$, we can divide the modified subproblem (\ref{SP_mod}) into two primal subproblems. The first primal subproblem can be defined as follows:
\begin{subequations} 
\label{Primal_1_SP}
\begin{flalign}
\underset{\boldsymbol{{y}}, \boldsymbol{\tau}_i}{\text{max}} \quad &  R(\boldsymbol{{y}}^{*j}, \boldsymbol{\tau}_i), \label{SP_OBJ_MOD_1}\\
\text{s.t.}  \quad   &   R(\boldsymbol{{y}}^{*j}, \boldsymbol{\tau}_i) = \sum_{i=1}^{M_h} z_i \Big( (1-{y}_i)R^{\mathrm{Local}}_{i} + {y}_i R^{\mathrm{LEO}}_i \Big), \label{SP_MOD_1_C1}\\
                     &   \sum_{i=1}^{M_h} \tau_i \leq T-\theta:~\lambda_1,  \label{SP_MOD_1_C2}\\  
                     &       \tau_i \geq 0, \quad \forall{i} \in   \mathcal{M}_h,\label{SP_MOD_1_C3} \\
                     &       {y}_{i} = y_{i}^{*j}:~\kappa_{i}^j, \quad \forall{i} \in   \mathcal{M}_h, \label{SP_MOD_1_C4} 
\end{flalign}
\end{subequations}
where $\boldsymbol{{y}^{*j}}$ represents the fix value of task computation decisions from master problem and $\lambda_1$ represents the dual variable associated with constraint (\ref{SP_MOD_1_C2}) in iteration $j$. The second primal-subproblem will be:
\begin{subequations} 
\label{Primal_2_SP}
\begin{flalign}
\underset{\tau_u}{\text{max}} \quad &  R(\tau_u), \label{SP_OBJ_MOD_2}\\
\text{s.t.}  \quad   &  R(\tau_u) = R^{\mathrm{UAV}}, \label{SP_MOD_2_C1}\\
                     &  \tau_u \leq \theta:~\lambda_2,  \label{SP_MOD_2_C2} \\
                     &  \tau_u > 0,\label{SP_MOD_2_C3} 
\end{flalign}
\end{subequations}
where $\lambda_2$ represents the dual variable associated with constraint (\ref{SP_MOD_2_C2}). After getting the optimal solution from both primal-subproblems, we need to update the fixed auxiliary variable value by the following equation:
\begin{equation}
    \theta = \theta - \zeta (\lambda_2 - \lambda_1),
\end{equation}
where $\zeta$ is a step size. Both primal-subproblems, i.e., (\ref{Primal_1_SP}) and (\ref{Primal_2_SP}) can satisfy the conditions of linear programming. Therefore, we can use a standard optimization solver, e.g., Gurobi \cite{gurobi}, to find an optimal solution. The details of the primal decomposition are summarized in the Algorithm \ref{PD}.
\begin{algorithm}[t]
	\caption{Optimal-Time Resource Allocation by Primal Decomposition} \label{PD}
	\begin{algorithmic}[1]
			\State Initialize: $\theta$
			\Repeat
			\State Solve the primal-subproblems in parallel
			\State \indent Solve problem (\ref{Primal_1_SP}) and acquire the optimal time-\indent \hskip1.5em resource allocation $\boldsymbol{\tau}^*_i$ for associated HUEs and \indent \hskip1.5em dual variable associated with constraint (\ref{SP_MOD_1_C2}).
			\State \indent Solve problem (\ref{Primal_2_SP}) and acquire the optimal time-\indent \hskip1.5em resource allocation $\tau^*_u$ for UAV-MEC and  dual \indent \hskip1.5em variable associated with constraint (\ref{SP_MOD_2_C2}).
			\State Update the time resource allocation auxiliary variable: 
			\indent \hskip1.5em $\theta = \theta - \zeta (\lambda_2 - \lambda_1)$.
			\Until{convergence}
	\end{algorithmic}
\end{algorithm}
\section{Simulation Results and Discussion}
\label{simulation_results}
\begin{figure}[t] 
\centering
{\includegraphics[width=\columnwidth, height=2.5in]{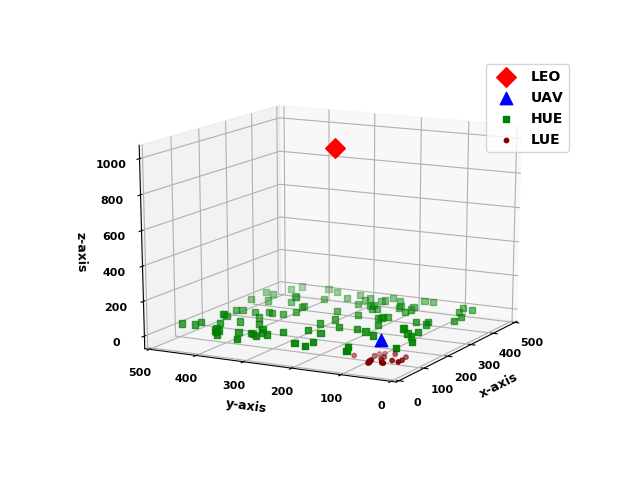}
\caption{Network topology consisting of LEO, UAV, HUEs
and LUEs.}
\label{network_topology}}
\end{figure}

For our simulations, we consider the HUEs in SAS-NTN to be uniformly distributed in $500$ nautical mile square area (NM$^2$) as shown in Fig. \ref{network_topology}. In this system, we deploy $M_h = 100$ HUEs within the coverage region of an LEO-MEC and CBS. Each HUE transmits power is considered as $P=33~$dBm because each HUE is assumed to be an energy-efficient device. The initial value of the lower bound to avoid infeasibility is set to $\Psi^{\mathrm{down}}$=$-25$. We use the python environment and the Gurobi optimization solver for implementing the proposed algorithm. Statistical results are averaged over 100 independent runs of random locations of HUEs. Our main simulation parameters are given in Table \ref{sim_tab}.

\begin{figure}[t] 
\centering
{\includegraphics[width=\columnwidth, height=2.35in]{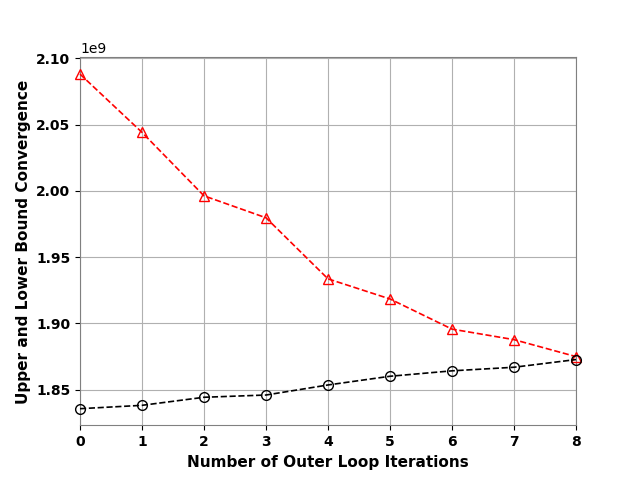}
\caption{Convergence of Bender decomposition algorithm.}
\label{convergence_BD}}
\end{figure}

In Fig. \ref{convergence_BD}, we present the convergence of the BD algorithm. 
The difference between the upper bound $R_{\mathrm{UB}}$ and lower bound $R_{\mathrm{LB}}$ value is decreasing and then finally converges. Fig. \ref{convergence_BD} shows that the BD algorithm converges to the optimal solution just after eight iterations, which is a very efficient convergence time.

\begin{figure}[t] 
\centering
{\includegraphics[width=\columnwidth, height=2.35in]{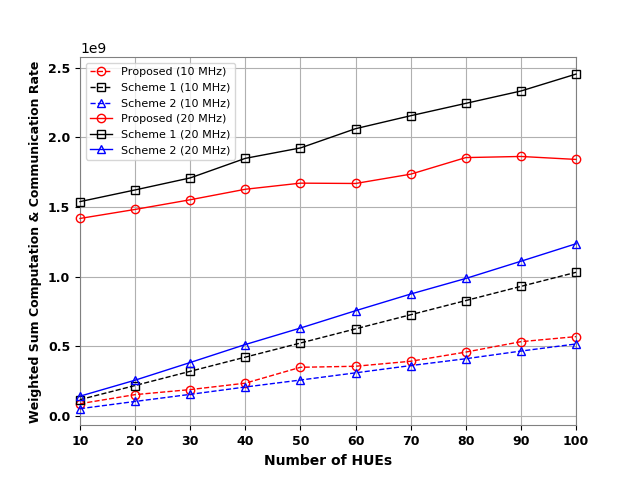}
\caption{Comparison of proposed algorithm with other schemes.}
\label{comparision}}
\end{figure}

In Fig. \ref{comparision}, the total weighted sum of the computation and communication rates versus the number of HUEs is presented. Fig. \ref{comparision} also demonstrates the comparison of the proposed algorithm with two schemes. Scheme 1 is considered as optimal results, which are computed by use of a standard optimization solver. Scheme 2 is regarded as a random task decision and time allocation to each HUE. Compared with scheme 2, our proposed algorithm performs better, which is close to optimal in the case of fewer HUEs. These patterns can be analyzed with both network bandwidth cases, i.e., $B=10~$MHz and $B=20~$MHz.
\setlength{\arrayrulewidth}{0.2mm}
\setlength{\tabcolsep}{6pt}
\renewcommand{\arraystretch}{1.1}
\begin{table}[t]
	\centering
	\caption{Simulation Parameters}
	\label{sim_tab}
	\begin{tabular}{|c|c|}
			\hline
			\textbf{Parameters}&\textbf{Values}\\ \hline
			Transmit Power & $P~$=$~33~$dBm  \\ \hline
			Noise Power & $\sigma^2~$=$~-104~$dB  \\ \hline
			Carrier Frequency  & $f~$=$~30~$GHz \\ \hline
			System Bandwidth & $B~$=$~20~$MHz  \\ \hline
			Communication Packet Overhead & $\mu~$=$~1.1$  \\ \hline
			Processor Cycles for one bit & $\chi~$=$~100~$  \\ \hline
			HUE Antenna Gain & $G_i~$=$~25~$dBi  \\ \hline
			UAV Antenna Gain & $G_u~$=$~25~$dBi  \\ \hline
            Satellite Antenna Gain & $G_s~$=$~30~$dBi \\ \hline
            Standard deviation & $\omega~$=$~0.1$ \\ \hline
            reference distance pathloss & $\Tilde{\gamma}~$=$~46.4$ \\ \hline
            pathloss exponent & ${\gamma}~$=$~2$ \\ \hline
            Rician fading parameter & ${\alpha}~$=$~1.59$ \\ \hline
		\end{tabular}
	\end{table}
\begin{figure}[t] 
\centering
{\includegraphics[width=\columnwidth, height=2.35in]{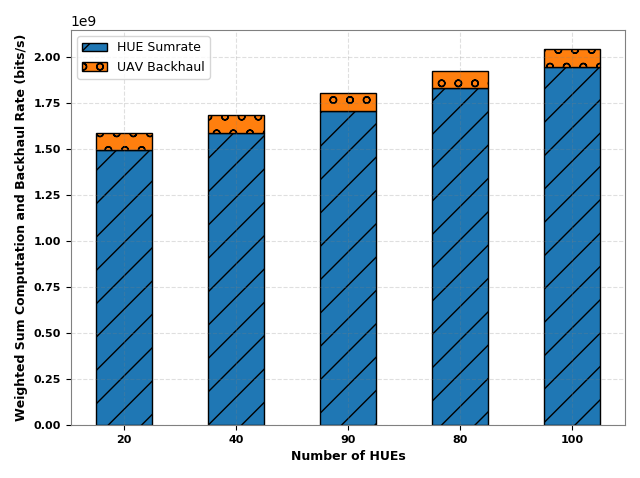}
\caption{Weighted sumrate (bits/s) vs HUEs.}
\label{data_comp_vs_users_VS_uav}}
\end{figure}

In Fig. \ref{data_comp_vs_users_VS_uav}, we present the total weighted sum of the communication, computation, and UAV-MEC backhaul rates. In this configuration, the system bandwidth is set to $B=20~$MHz. We can observe that, in each scenario, the network allocates time for HUEs computation and UAV-MEC backhauling. As the number of HUEs increases, the network performance improves because more HUE needs to execute their task at the LEO-MEC satellite. This LEO-MEC enabled the SAS-NTN network to perform well for both HUEs and UAV-MEC backhaul demands.
\section{Conclusion}
\label{conclusion}
In this paper, we devised task decisions and time resource allocation for the blue data computation maximization in space-air-sea non-terrestrial networks. The proposed problem is designed to be mixed-integer linear programming. To solve this problem, we have proposed a joint Bender and primal decomposition algorithm. Bender decomposition is an outer structure algorithm that divides the initial problem into master and subproblem and then solves iteratively. To deal with the coupling constraint in the subproblem, we proposed a primal decomposition algorithm to solve the subproblem iteratively. Simulation results have shown that the proposed algorithm can yield better results with efficient convergence time. In our future work, we will consider the user's mobility in the network.
\bibliographystyle{IEEEtran}
\bibliography{GLOBECOM}
\end{document}